\documentclass[showpacs,superscriptaddress, preprintnumbers,amsmath,amssymb]{revtex4}

\usepackage{graphicx}
\usepackage{dcolumn}
\usepackage{bm}
\include{epsf}

\begin{document}

\preprint{CHESS 09}

\title{On religion and language evolutions seen through mathematical and agent based models}

\author{M. Ausloos}\email{marcel.ausloos@ulg.ac.be}
\affiliation{GRAPES@SUPRATECS, ULG, B5a Sart-Tilman, B-4000 Li\`ege, Euroland
}


\date{today}

\begin{abstract}
Religions and languages are social variables, like age, sex, wealth or political opinions, to be studied like any other organizational parameter. In fact, religiosity is one of the most important sociological aspects of populations.  Languages are also  obvious characteristics of the human species. Religions,   languages appear though also disappear.  All religions and languages evolve  and survive when they adapt to the  society developments. On the other hand, the number of adherents of a given religion, or the number of persons speaking a language is not fixed in time, - nor space.

 Several questions can be raised. E.g. from a macroscopic point of view :  How many religions/languages exist at a given time? What is their distribution? What is their life time?  How do they evolve? From a ''microscopic'' view point:  can one invent agent based models to describe macroscopic aspects? Does it exist simple evolution equations? How complicated must be a model?
 
These aspects are considered in the present note.   Basic evolution equations are outlined and critically,  though briefly, discussed. Similarities and differences between religions and languages are summarized.  Cases can be illustrated with historical facts and data. It is stressed that characteristic time scales are different. It is emphasized that ''external fields'' are historically very relevant in the case of religions, rending the study more ''interesting'' within a mechanistic approach based on parity and symmetry of clusters concepts.  Yet the modern description of human societies through networks  in reported simulations  is still lacking some mandatory ingredients, i.e. the non scalar nature of  the nodes, and the non binary aspects of nodes and links, though for the latter this is already often taken into account, including directions. 

From an analytical point of view one can consider a population independently of the others. It is intuitively accepted, but also found  from the statistical  analysis of the frequency distribution that  an attachment process is the primary cause of the distribution evolution  in the number of adepts: usually the initial religion/language  is that of the mother. However later on, changes  can occur either due to ''heterogeneous agent interaction'' processes or due to ''external field'' constraints, - or both.  In so doing one has to consider competition-like processes, in a general environment with different rates of reproduction. More general equations are thus proposed for future work.
\end{abstract}



\maketitle

\section{Introduction}\label{intro}

Language and religion like sex, age, wealth, job, political affiliation,  ... can be considered to characterize a human individual  in its community \cite{chomsky,durkheim}. Other characteristics may serve to  identify a human being, like through its music interests \cite{music1,music2}  or  its previous or present health conditions; see typical questions "who are you? " on visa application forms. These are all examples of some sort of  ''degrees of freedom'' as so often found in statistical physics of critical phenomena \cite{hes}  and dissipative structures \cite{prigogine}, self-organizing or not, in order to describe phase   symmetries. Thus  statistical physics ideas can find an interesting role outside physics  \cite{stauffer04,carbone}  not only in econophysics but also in research pertaining to attitudes, behaviors, opinion formations, etc., as in   socio-economic studies. In particular investigations on religion and language distributions, on one hand,   and their evolution, on the other hand, can be studied as if such a degree of freedom is a variable defining some thermodynamic state, as a function of time and space \cite{mimkes}. These  so called microscopic characteristics, can be studied  through auto-correlations, or correlated with any other variable or ''parameter'' characterizing a population made of so called agents.  The results can be next compared with available global data on macroscales.

Several pertinent questions can be raised, e.g.  
from a ''macroscopic'' point of view :  (i) How many religions/languages exist at a
given time? (ii) How are they spatially distributed ? ... From a ''microscopic'' view point:  (iii) How many 
persons  consider that they practice one religion/language?  (iv) Does the number of adherents
increase or not, (v) and how?  and maybe why?  
(vi) Thus what are, if any, the basic equations ? 
(vii) Last but not least is there some possible modelization, through some agent based model?

However in this limited size contribution it is impossible to present a full state of the art concerning present studies by sociophysicists on languages and/or religions. Only a few topics can be tackled. The selection is admittedly biased and mainly due to a serious lack of competence of the author  in front of huge research field activities. In Sec. \ref{sim_n_diff} the differences and similarities between religions and languages are emphasized,  with some stress put on the difference in time scales and the external field constraints. In Sec. \ref{basiceqs},   basic equations are written in to illustrate fundamental approaches to religion and/or language evolutions, without going into details. Many reviews exist elsewhere.

 Sec. \ref{nets}  serves to indicate that human networks are more complex than those studied  by sociophysicists presently performing simulations of agent based models. In particular the notion of node intensity should appear to be relevant in view of the considerations in the previous sections.
In Sec. \ref{finaleqs} a more general set of equations is thereafter  proposed and illustrated through an example involving the competitions between a few ''degrees of freedom'' of different types, like if there are  describing struggles between ideologies, in a general sense. A conclusion  and suggestions for future work  follow
in Sec. \ref{disc_n_concl}

\section{Similarities and differences}\label{sim_n_diff}

   Through this      section, it is hoped  to outline  a few   ''differences'',  also recognizing similarities, between languages and religions, from both an anthropological and  a physics point of view. This is summarized through  several items in  Table \ref{tab01AppA} which are  briefly commented upon  \cite{footnotetable}. This should serve as a perspective or input into modeling their sociological features in Sec. \ref{basiceqs} and Sec. \ref{finaleqs}. 
        
        For example,  the $origin$ of languages and/or religions cannot be easily discussed as a simple statistical physics process. Basic  anthropological studies, like those of  refs.\cite{Ruhlen,Johansson,Diamond,Campbell,Nichols,Greenberg}, or  refs. \cite{eliade,swanson,dubuisson,Dennett}   are sometimes considered as hypotheses or theories.   It is accepted that languages could not have originated as they are without a physiological modification of the human animal. Moreover, diversification is not only due to events like the Babel Tower construction, but  is also rather due to different environmental, economic  and  sociological conditions.  Similarly, religions could not originate from myths subsequent to the ''scientifically unexplained'' observation of natural phenomena. However the ''final result'', i.e. the appearance of a language or a religion \cite{upal} can be imagined by a physicist to correspond to a nucleation/growth process through homogeneous or heterogeneous fluctuations, {\it exordium} to competiton between surface and volume free energy terms. It is relevant to point out to the attempt by Sznajd-Weron to demonstrate how to predict the initial concentration necessary for the fluctuation to grow to a viable size \cite{KSW} in an opinion model of societies. Nevertheless it will take some time before one can connect such (microscopic) fluctuations  (... of what ?) to the truly anthropological approaches. Always containing very  profound and  elaborate thinking. 
        
           First let it be recognized that the definition of a religion  or an adherent (or adept) might not be accepted univocally, but the same can be said about languages. It is much debated between scholars on how to define  a religion or a language.  Like many, one can put on the same
footing religions, philosophies, sects and rituals \cite{Dennett}.  It is accepted, e.g.,  that a sect is a legally defined entity, - which in fact depends on the country where the $legal$ definition is made.
 $Idem$,  one could distinguish between adherents or
adepts;  there are also agnostics, atheists or ''not concerned''. 
Annoyingly there are various denominations for ''closely related'' religions which can impair data
gathering and subsequent analysis.  A similar set of considerations exists when discussing languages and
dialects, slangs, etc.: e.g. there are three practical definitions of a  language in \cite{klinkenberg}. The variety of religions and languages  is surely enormous, to the point that they may even appear as fully individualistic ones, and only  gathered into bins for defining communities.

It is then a habit of human beings to find roots and to connect to some important feature or on the contrary be singled out. Thus finding some genealogical-like trees for languages through their vocabulary or/and grammar, e.g. see \cite{petroniserva1,petroniserva2,zanette_l,Wichmann_text}, or similarly for religions, e.g. see \cite{religioustree} can be done through various historical facts or measures.   Such  work on so called language trees is done like in many approaches about hierarchical systems.   Yet it seems forbidden at this level and to-day  to rank religions through any other indicators than materialistic ones, i.e. having led to some quantitative measure,  - in contrast to written texts in some language, including translations, for which it is more common to discuss  their  ''quality'  or ''values''. One should be careful at another level:  take the example  pointed out by Nash \cite{Nash7} in the case of christianity. He recalls that in searching for connections with other beliefs, scholars who first use  a given  terminology to describe e.g. pagan beliefs and practices,  then marvel at the awesome  parallels they think they have discovered. See also \cite{mythology}.  
Nevertheless although it might be an audacious  suggestion to consider the hierarchy of languages and religions when describing them through network schemes  (see below) , it might be considered to use a  measure, based on concepts,  as the average overlap index (AOI)  for  some ranking \cite{gligor512,gligor537,gligorAOI},  as done for countries in econophysics studies. This jhas led to some original observation about communities.


It should be stressed that statistical physics needs data in order to produce some thought and answer a few questions. See the difficulty of obtaining and analyzing  data in sociology studies.  The data which is often used for describing the language distribution comes from \cite{WALS,Grimes}; that for religion has been taken from \cite{WCT,WCE}.
However there are many other places where some data can be obtained, including  detailed geographical ones \cite{UScensus_r,UScensus_l}.  Interestingly one can also connect to data obtained taking into account  various economic aspects of religions \cite{ianna98}, maybe not so easily about  the economics of languages.

Interestingly it is found, by survey takers, that there are about 6000 languages and 3000 religions. One may wonder why the order of magnitude is similar. The more so when it is absolutely clear that one can be a polyglot, but it is $very$ difficult to be polyreligious, - even though the caveat pointed out here about holds much strongly here: religion  being sometimes $very$ individualistic. Therefore the log-normal plots of the {\it probability  distribution function} (PDF) of the number of practitioners of a language \cite{wichman,paulo3} or a religion \cite{religion1} are very strikingly similar, including on the value of the large tail, i.e. respectively   -1.42 \cite{paulo} and -  1.67  \cite{zanette_r}.
In both cases, the effect of the
  enhanced number of languages for very small sizes is  nevertheless very perturbing, in particular since it announces the forever  disappearance of  rare cases. 

A critical view of this data has to follow:  in \cite{religion1}   a break was noticed in the PDF  at $10^7$ adherents,   indicating  
an overestimation of adepts/adherents  in or by) the most prominent religions, or a lack of
distinctions between denominations, for these, - as can be easily understood either in terms of propaganda or politics, or because of the difficulty of surveying such cases precisely. Yet one paradoxical surprise stems in the apparent  precision of the data. E.g., in several cases,  the data in \cite{WCT} seems to be precise up to the last digit  i.e.,  in mid-2000 , there are  1057328093 and 38977 roman catholics and mandeans respectively.  In strong contrast there are 7000000 and  1650000 wahhabites and black muslims respectively, numbers which are quite well rounded. Thus a mere reading of the numbers  warns about the difficulty of fully trusting the data. Nevertheless the analysis should be pursued bearing this $caveat$ in mind. One can here point out  an   application of  PDF studies on Jehowah witness activities, -   as if they were   economic ones \cite{PicoliMendes08}. 

After having examined the global aspects, let us turn toward more microscopic considerations, i.e. the agents themselves,  the role of opinion leaders and the external field effects on agents.

From this  time point of view,  and stressing the time scale, one can notice that a religion can seem to appear rather instantaneously, often as a so called sect,  at the beginning, and its number of adherent can grow steadily (see the recent Mormon or Rastafarianism case) or not.  New religions ,  necessarily called sects at their beginning, appeared after the second world war in Japan. Many disappeared. But in other cases, like the  Black muslims, Bahai,  Mormons, Sokka-Gakkai,
Universal Church of the Kingdom of God (founded in 1977, it has already more than 2 millions adepts), 
Jehowah witnesses,  the
Scientology church, etc.   New languages can also appear, for very specific reasons: let Esperanto,   FORTRAN and other computer ''languages"", be mentioned here \cite{fortran,esperanto}. The notion of nucleation through homogeneous or heterogeneous fluctuations, in some imposed (thus external) field,  could be thought of.  A religion can also  rather quickly disappear \cite{antoinism}, like the Antoinists in some coal mine regions of Western Europe,  because its social role or its feeding ground  disappears.  Both cases  though quite interesting are actually outside the realm of this paper. Yet the  life time, or aging, of a religion can be studied, through the number of adherents, surely for modern times.

 Notice the interesting fact when strong fluctuations arise: history is full of examples of individuals or entire groups of people
 changing their religion, - for various reasons: following the ''leader'', e.g. Constantinus,  Clovis, or not changing at all under ''external pressure'' , leading to martyrdom, or  like at inquisition time, or following a fatwah,  but also cases of  ''internal pressure'' (Khazars \cite{khazars}, maybe) or so called adaptation under proselytism action, e.g. sun worshipping Incas in presence of catholic missionaries,  Zoroastrians in  Persia  and Bogomils in Bosnia in presence of Muslim Arabs, etc. Such a competition through agent interactions or under ''external field conditions'' exist in many religion cases indeed. Thus, the number of adherents can  evolve drastically due to such various conditions \cite{roach1,roach3}, independently of the population growth size. Notice that it can be also a ''legal field'' which decides upon the use of a language. Fran\c{c}ois I, King of France, decided in Villers-Coteret in 1529 that  the true french was the one  $spoken$ in Isle of France, and nothing else.  Back to religion history, a quarter century 
later, after the Peace of Augsburg treaty (1555), the ruling prince could decide in Germany whether
his territory is protestant or catholic: {\it Cuius regio, eius religio}.   In all cases,   that does not mean that everybody became a christian in the Roman Empire nor everybody spoke french in what was France, at that time,   but it could lead to intolerance, wars of religions,  and massacres.   
One can recall the case of ''Matines Brugeoises'' or ''Brugse Metten'' (1302)  when flemish ''peasants'' killed the french nobles, recognized as such, because they could not pronounce correctly "schild and vriend''. Other cases are that of the red khmers killing Vietnam educated intellectuals in Cambodia and that of Gileadites killing Ephraimites at a Jordan ford. 
Yet it is hard to find massacres due to ''language heresy'', though discrimination is known due to language conflicts, - see the problems of the French speaking community around Brussels, Belgium !   or the contempt of european portugese  for the brazilian portugese; kurd and armenian languages in Turkey; not too long ago, french in Canada, etc.   

However religious (so called) heresy and differences has led to many more deaths:  no need to mention  Iran,  China,  Palestine, Balkan. This might be due to the fact that religions can serve as ground or excuses for political union: see Poland under the communist regime. Notice that attempts to suppress some ''heresy'' through harsh means does not always work: see the rise of protestantism in Great Britain even through strong  means \cite{roach2}, though inquisition had some effect  \cite{roach1,roach3}.  

 See also a mapping of a sometimes called (religious or scientific) controversy on networks made of Intelligent Design Proponents (IDP) and Darwinian Evolution Defenders (DED)  in order to analyze the local and global structural properties \cite{garcia} of such strongly diverging opinion  networks.  See also the approach toward linguistics within the network concepts, not necessarily involving agents, in \cite{bif5,DorogovtsevMendes01,ke}.
 
  Yet, it  has been found in \cite{religion1}  that empirical laws can be deduced for the number of adherents, i.e. the {\it probability distribution function}.  Two quite different statistical models were proposed, both reproducing well the data, with the same precision, one being a preferential attachment model  \cite{prefatt},  like for  heterogeneous interacting agents on evolving networks, e.g. it is more likely that one has the religion of one's mother or neighbor..... (leading to a log-normal distribution), another based on a ''time of failure'' argument (leading to a Weibull distribution function).

          \begin{table}

\caption{Comparing   languages and religions through a few similarities and differences}
\smallskip
\begin{footnotesize}
\begin{center}
\begin{tabular}{|c| c| c c c|  c c   c|c| }
\hline 
 & \multicolumn{4}{|c|}{Languages 	}& \multicolumn{4}{c|}{Religions 	}\\ \cline{2-9}
 		&Refs.&     	&	 	&		 &		 	&	 &	&Refs.
\\\hline
origin	&\cite{Ruhlen,Diamond,Johansson}&   physiology& ''Babel Tower''	&&natural phenomena		 &	myths	 	 &&\cite{durkheim,eliade,swanson,dubuisson}	
\\\hline
 variety	 &\cite{Campbell,klinkenberg}& 	huge: 		& dialects, slangs&  	&  huge: 	& denominations, sects 	&&		 
\\\hline
semantics 	&\cite{Greenberg,klinkenberg}&	grammar	& vocabulary &	  	& images		& rituals  	&	 	&\cite{Dennett}
\\\hline
agents 	&\cite{Campbell,Nichols}& multilingual 	& frequent &&	polyreligious 		& rare &&		 
\\\hline
impact factor & &citations&	 libraries&	  &''saints''& worship sites&	 &
\\\hline
data set	&\cite{WALS,Grimes}&  & WALS &	 	&		 &		WCT, WCE 	& 	&	\cite{WCT,WCE}
\\\hline
number 	&&  & more than 6000 &	 	&		 &		more than 3000 	& 	&	
\\\hline
PDF	&\cite{Schulzereview07}&  &$ \sim$   log-normal &	 	&		 &		$\sim$  log-normal 	& 	&	\cite{religion1}
\\\hline
PDF	tail&\cite{wichman,paulo}&  & -. 1.42 &	 	&		 &		- 1.67 & &\cite{zanette_r,religion1}	 	
\\\hline
genealogy	&\cite{petroniserva1,petroniserva2,Wichmann_text}&  & tree &	 	&		 &		tree	&&\cite{religioustree}	
\\\hline
hierarchy	&&  &(AOI ?) &	 	&		 &		(AOI !) 	& 	&	
\\\hline
opinion leaders	& & authors	& teachers &&	 (high) priests & witches, shamans&&		 
\\\hline
time scales 	&& 			&		&		&	 		&	 	&				&		\\   && nucleation  	& slow  	&		& nucleation	& fast 	&	  &
		\\	&& slow growth  	&  Fran\c{c}ois I&	  & Contantine, Kazars& fast 	through	avalanches&&  		 
		\\	&\cite{NowakKrakauerPNAS99,Abr+03}& decay 	&  &&	decay	& &	  &\cite{antoinism}		
\\\hline
applied fields	&& 			&		&		&	 		&	 	&				&				\\	&& Brugse Metten&discrimination	&& inquisition  &Bogomils, Cathars&	&
	\\	&&competition	& media	&	  	 	& competition& economics	&  &  	 \cite{roach3,ianna98}	
		\\   && enforced civilization & ... &		&proselytism  	& often, strong	&	  &
		 
\\	&&  &minorities	&&  & decay, revival & 	&
\\\hline
network&\cite{bif5,DorogovtsevMendes01,ke}&   small world &&	 	&&IDP-DED		 	& 	&	\cite{garcia}
\\\hline
\end{tabular}
\label{tab01AppA} 
\end{center}
\end{footnotesize}
\end{table}

\section{Basic equations of evolution}\label{basiceqs}

The most fundamentally relevant variables are thus accepted to be the number of practitioners of a language/religion  \cite{footnote2} Only this number is treated here  as the physics object/measure.
Thus a language/religion is hereby considered as a (socially based)
variable to be so studied like any other
organizational parameter for defining a thermodynamic-like state.  Since a religious state \cite{footnote3}  

is more individualistic than a linguistic state, one can better define the religious adherence of an agent than the linguistic one.   Indeed one can hardly be multi-religious but one can be a polyglot. Of course one can switch  easily, i.e. through ''conversion'',  from one religious denomination to another; this is not so easy in language cases, - except through miracles. Thus the observation time of a religious state needs careful attention in surveys, but is less so drastic for language studies, as already hinted here above.

The dynamics of world's languages, especially   their disappearance  \cite{NowakKrakauerPNAS99,Abr+03} has been  recently considered through a Verhulst  time $t$ evolution   equation
\cite{verhulst} for the number density $\rho_i$ of practitioners of some language $i$

\begin{equation}\label{verhulst}
\frac{\partial \rho_i}{\partial t} = r_i\; \rho_i \left( 1- \frac{\rho_i}{C_i} \right)
\end{equation}
where $C_i$
is the carrying capacity of the environment  for the population speaking the $i$ language
and $r_i$
is a positive or negative growth rate.

It was soon discussed that the evolution of one language must take into account some competition with other languages  
\cite{Viviane,DorogovtsevMendes01,ke,ThelwallPrice06,staufferreview,Schulzereview06,Schulzereview07}, but also geographical constraints \cite{SchulzeStauffer07} and social structures \cite{MinettWang07}, sometimes extending the concepts toward hard evolutionist views \cite{hashemi,steele2}.  Whence the Verhulst equation must be supplemented by a diffusion equation

\begin{equation}\label{diffusionvec}
\frac{\partial \rho_{i}}{\partial t} + {\rm div} \vec{j}_{i} = C_{i}
\end{equation}

where  $\vec{j}_{i}$ describes the net flux of the population  speaking the language $i$ into some area 
or in terms of a diffusion coefficient along  some $x$ axis 

\begin{equation}\label{diffusionD}
\frac{\partial \rho_{i}}{\partial t}   =   \; D_{i} \; \frac{\partial^2 \rho_{i}}{\partial x^2} .  
\end{equation}

There are several review papers like \cite{staufferreview,Schulzereview06,Schulzereview07} discussing such cases. 

A set of similar equations for religions can be  written when attempting to quantify some religion dynamics  \cite{futures,hayw99,hayw05,shy,religion1} through the number of adepts. 
A population growth-death equation has been conjectured  indeed to be a plausible modeling of  such religion 
evolution dynamics, in a continuous time framework. The time evolution of several ''main'' religions was considered to be described, at  a so called microscopic level,    along the lines of
the  Avrami-Kolmogorov equation describing solid state formation in a continuous time framework \cite{auslooscrystalgrowth}. The solution of which is usually written as

\begin{equation}\label{Avrami}
f(t)=1- e^{- K t^{n} }
\end{equation}
where $f(t)$ is the volume fraction being transformed from one phase to another; $K$  and $n$ are adjustable parameters.
  
 {\it A priori} in analogy with crystal growth studies \cite{auslooscrystalgrowth,Gadom}, we have considered that a microscopic-like, continuous time differential equation can be written for the evolution of the number of adherents, in terms of the percentage with respect to the world population,  of the world main religions, as for competing phase entities in Avrami sense
 
\begin{equation}\label{Avramidiff}
{d \over dt} g(t)=\gamma t^{-h} [1-g(t)].
\end{equation}
It can be solved easily giving the evolution equation for the fraction $g(t)$ of religion adherents  
in terms of  a (positive for growth process) rate (or scaling) parameter to be determined,  and   a parameter  $h$ measuring the attachment-growth (or death) process to be
deduced in each case, 
in a continuous time approximation.  It    should be emphasized  that religions have appeared at some time $t_0$ which is somewhat unknown, or to say the least, loosely defined, due to a lack of historical facts but also due to the inherent process of the creation of a religion. 
In both, religion and language, cases, the initial condition  (IC) which must be imposed in resolving the above equations is a major difficulty. Fortunately one might have some measure of $\rho_i$ at some given time before integrating.

 Thus to address some of these issues,  classical scientific steps can be followed as in
physics investigations \cite{religion1,religion2,religion3}.   $Accepting$ as  valid the number
of adherents of religions from various surveys,  ''empirical'' data can be subsequently analyzed.  
Examples of religions  were found for which the number of adherents is
increasing (e.g., Islam), decaying (e.g., Ethnoreligions) or
rather stable (e.g.,  Christianity and Buddhism), - all giving interesting values for $h$.  This has led to
a  more complete and   more detailed analysis of the values of $h$ and of  its meaning for 58 ''time series'' \cite{footnotedata}  in \cite{religion2,religion3}.
Anomalous, or unexpected $h$ values have been attributed to so called 
external field  conditions,  somewhat based on intuition.  

\begin{figure}
\centering
\includegraphics[height=10cm,width=10cm]{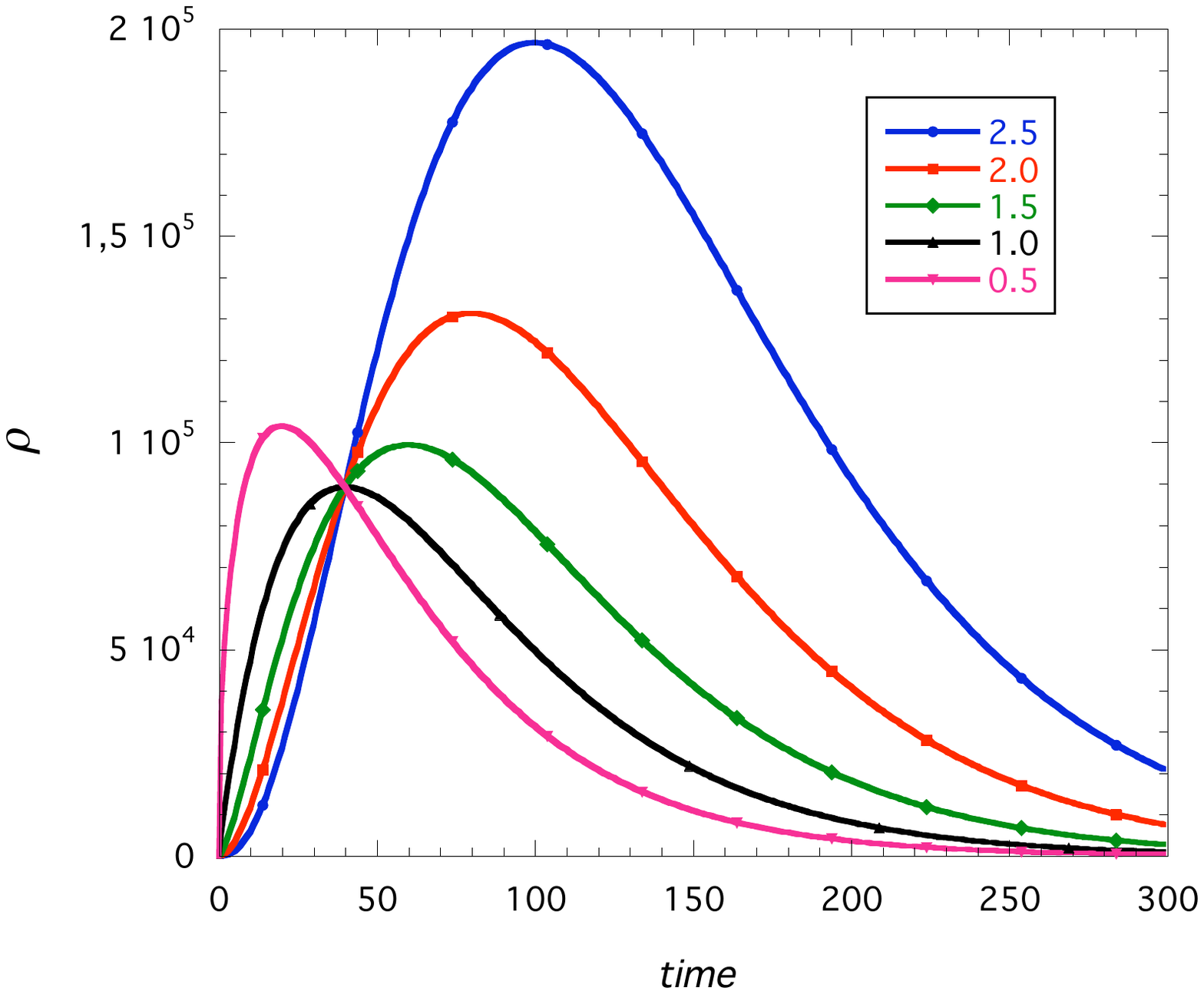}
\caption{\label{fig_antoine_white}   Graph representing the evolution in the number of adepts under grow/decay conditions, as in Eq.(\ref{antoineint}),  as  for the Antoinist. The $y$- and $x$- axis  values are arbitrary; for the display $t_1=40$ and $\alpha= 20$ values were chosen; the value of the $b$ exponent is indicated in the inset}
\end{figure}

However the main criticism of these   equations  is that  they do not fully take into account social and historical conditions, - in some  wide sense, the external/political fields and the environment, in a loose sense. For example in the case of the Antoinist which was founded by Louis Antoine at the end of the XIX-th century in some Li\`ege suburb and had a number of adherents up to about 200 000,  after 50 years of existence or so, and nowadays decays, it is found that the evolution equation is better represented  through a rate equation
which takes into account an external (sociological or not) condition through a parameter, $c$, i.e.

\begin{equation}\label{antoine}
\frac{d \rho}{d t} =  b\;  \left(   \frac{\rho}{t} \right) - c\; \rho
\end{equation}

such that  the evolution of $\rho$ reads

\begin{equation}\label{antoineint_0}
 \rho( t) =  a\;    \left(   \frac{t}{t_1} \right)  ^b \;  e^{-ct}
 \end{equation}
where $a$ is an integration constant which can be adjusted. This equation states that the relative rate of change of the numbers of adherents $\frac{d\rho}{\rho} $ is proportional to $\frac{dt}{t} $  instead of to
$\left( 1- \frac{\rho}{C} \right) \; dt $.  and a scaling time $t_1$.  For a proper display   (Fig.\ref{fig_antoine_white}) a change of variables is made such that the previous equation reads
\begin{equation}\label{antoineint}
 \rho( t) =  A\;     \left(   \frac{t}{t_1} \right)  ^b  \;  e^{-  \frac{t}{t_1} +\alpha}
 \end{equation}

Other formulations can be found that  would present such a maximum after an appropriate time; they would need to be investigated and interpreted.

 In concluding this section it is fair to point out that the spreading of languages and religions can be mapped into some  approach like epidemics modeling \cite{roach3}. At the same  time, it can be thought that the spreading is similar to market sharing. Indeed the conclusions of Roach \cite{roach3} when distinguishing between the behaviors of Cathar heresy in France and that of Bogomils in Bulgaria can be mapped either as an epidemic  control  or a market share battle depending on whether or not one looks at the overall  result or at the cause of the result. 
 Other spreading taking into account asymmetric grow and environmental conditions,  can be found in anthropological research, see e.g. Paleoindian spreading in \cite{steele}. In most of these cases, the grow is analogous  to crystal growth \cite{iop,candiareview,VdWMA_PRL,167,187,205}  conditions.

\section{Networks}\label{nets}

\subsection{Cluster expansion}   

In  physics one should study the response of the system to intrinsic or extrinsic fields.    In a mechanistical approach, the population of agents through a free energy $F$, Hamiltonian  formalism or Langevin equation. In such a way, one would develop the quantity of interest as a series in terms of clusters, e.g., ordered along  the increasing size of the cluster according to the number of spins  in the cluster $<...>$,  as in
  
\begin{equation}\label{A1}
exp\left[-\frac{F-F_0}{kT}\right] = - \Sigma_i H_i S_i -   \Sigma_{<ij>}  J_{ij} S_i.S_j-  \Sigma_{<ijk>}  K_{ijk} S_i.S_j.S_k - \cdots  \;\;\; , \;\;\;
  \end{equation}
  
 in obvious notations, i.e. each spin $S_i$ representing an adept  in an external field $H_i$ and interacting with another adept through some  interaction $J_{ij}$, etc.,  or similarly
  
\begin{equation}\label{A2}
{\partial \Phi_i \over \partial  t}=  A_{ij} \Phi_j+  B_{ijk} \Phi_j\Phi_k  +  C_{ijkl} \Phi_j\Phi_k\Phi_l\\  + \cdots  \;\;\; , \;\;\;  \end{equation}  

for some information flux $\Phi_i$. A vector generalization is immediately thought of.
    
   \subsection{Node intensity}  
   
     Although language is primarily $spoken$, most of the relevant research   is about its  $written$ forms of communication, - which have significantly different sets of properties (known as registers) (e.g., \cite{Biber2003}) than  spoken forms. It occurs that Web documents  e.g. come from a wide range of styles, while blogs probably tend  to be relatively informal, some being close to spoken language. In contrast, academic web sites contain large collections of very formal documents, such as research papers in e-journals, online copies of computer documentation, and university rules and regulations. Nevertheless, they also contain less formal genres such as personal home pages. Nevertheless one  should clearly  precise what is meant by "language investigations'', since one could  wonder whether someone pretending to be a language practitioner, indeed understands, reads, speaks,  writes, ... the language. Moreover it seems relevant to question the  quality level of the competence.
     
Similarly  the time dependence of the  $number$ of adherents can be considered to be a very restrictive, and even wrong,     way to ''measure''  the evolution of a religion.   One should also ''weight'' the level of adherence to a religion. For example, one could try as for languages to define a religion through its member intensity of participations in rituals, and in practicing the principles. Indeed, one could distinguish between $adepts$ and $adherents$. An adept definition can be found in \cite{wikiadept} 

 One can measure diverse quantities related to the  religious effect as well. 
                Moreover,   one could consider  religion dynamics from another ensemble of points of view or  $indicators$ \cite{Herteliu}.
As there are  several definitions of a  language \cite{klinkenberg}, similarly one could also define what a religion ''is'' in different ways  \cite{Dennett,religioustolerance,religioustolerance2}.

             Thus it is  clear that a religious/language ''adherent'' instead of being an analog of an up or down spin, is rather a vector for which each element can be a quantity measuring some $value$ like one of those considered  in sociology, i.e. a ''quality''.  The measure of this quality being controversial depends on whether it is a self-assesment, on whatever scale, or a direct measure of some type.  This make modelling very much complex.
             
             No need to say that physicists are not the first ones to reflect on variability in religion distribution  or adherence level. One may find already such considerations in  books and papers by specialists of the history or sociology of religions, e.g.  \cite{Dennett}.

 Therefore it is emphasized that one should describe  society through   agents having degrees of freedom which are both qualitative and quantitative in nature. Each node of the network becomes a quite complicated mathematical entity, surely not a scalar. 
  Each node can be better considered as a vector node, each
component being like  a degree of freedom,  represented by a tag, itself
having an intensity, like on some energy level in a molecule. Interactions
between the various degrees of freedom of different nodes can occur,
through a matrix form; each link having a weight and a direction. That
interaction matrix might not be symmetric. The more so if time delays are
taken into account. Should one say, like molecules made of ions, or nuclei
  made of elementary particles interacting in complex ways? Should one at this stage recall Haven \cite{Haven1,Haven2} description of economy and economic agents along quantum mechanical-like concepts? Indeed quantitative and qualitative dynamical evolutions of agents and groups (''denominations'') can   find some theoretical source in many competition and organization physics models.
Next, Potts vector or ferrroelectric type (Hamiltonian) models  can be imagined for describing an ensemble of religious  agent state or evolution.

Thereafter  but {\it exordium} it seems natural to go back to the classical  Verhulst ideas and its extensions for prey-predator problems in a  Lotka-Volterra model 
by introducing some
realistic conditions on the growth ratios and on the interaction coefficients  between  the agents in the
populations \cite{dv00,dv06}. - 
Indeed there is  an obvious analogy with the problem of extinction of populations \cite{pec08}, with those of
religions \cite{religion1,religion2,religion3} and languages \cite{Abr+03}, and also to the very modern  question of
internet governance 
\cite{olivera}.

    \section{Generalized model}\label{finaleqs}
    
  Leaving aside the networking description of the society, let us consider a set of Lotka-Volterra equations to describe the ''struggles between ideologies'', religions or languages, i.e. in a very general sense.  It has been proposed to consider the following scheme \cite{vitanov1,vitanov2}. 
    
 Let  us consider a set (''area'') with a population of $N$ agents.
Let the population   be divided into $n+1$ factions or communities:  each faction has  a 
 specific ideology,  such that  the
number of members in the corresponding community is $N_{1}, N_{2}, \dots, N_{n}$, but with a singled out 
fraction $N_{0}$ of people which are  considered to be $not$  concerned, not following  an  ideology, not speaking any language,  at a given  time.  By definition at each time $t$, 

\begin{equation}
N = N_{0} + \sum_{i=1}^{n} N_{i}
\end{equation} 

The overall population $N(t)$ is assumed to evolve according to the generalized Verhulst law

\begin{eqnarray}
\frac{dN}{dt}=r(t,N,N_{\nu}, p_{\mu}) N \times 
 \left[ 1 - \frac{N}{C(t,N, N_{\nu}, p_{\mu})} \right] 
\end{eqnarray}

where $N_{\nu}$  represents the set   $N_{1},\dots,N_{n}$ while 
 $p_{\mu}=(p_{1},\dots,p_{m})$ is a set of parameters describing the environment, 
and  $r$ is the overall population growth rate; which can be positive or negative.

    The growth process is constantly disrupted by small extinction events, as in \cite{wilke}, 
monitored  through $r(t, N, N_{\nu},  p_{\mu})$. As before $C(t,N,N_{\nu}, p_{\mu})$
 is the carrying capacity  in the area.
 
  \subsection{General remarks}

  In every community   $i$ one has to account for the  following processes:
deaths, dissatisfaction, unitary conversion, and binary conversion.

\begin{enumerate}
\item
First,    the number of  followers of   an  ideology can decrease through death or  dissatisfaction
with the ideology,  i.e. through a term $r_{i} N_{i}$, where $r_{i} \le 0$. 
In general $r_{i}=r_{i}(t,N,N_{\nu},p_{\mu})$.

\item
Conversion from one ideology to another 
can be made $without$ direct contact between the followers of different ideologies.
This type of conversion happens through the information environment of the population.
 In order to model the ''unitary''
conversion we assume that the number of people converted from ideology $j$ to ideology $i$
is proportional to the number $N_{j}$ of the followers of the ideology $j$.  A field 
$f_{ij}$  coefficient characterizes the intensity  with which this conversion  from $j$ to $i$ occurs. Observe that $f_{ij}$ is not necessarily a scalar.
The corresponding modeling term is $f_{ij} N_{j}$; $f_{ii}=0$. 
  In general
$$
f_{ij}=f_{ij}(t,N,N_{\nu}, p_{\mu},C)
$$ 
 
\item
A conversion to the
$i$-th ideology can occur because of direct interaction  between a member of this $i$ ideology with a member of some other $j$ ideology.
It can be at first assumed that the intensity of the interpersonal contacts is
proportional to the numbers $N_{j}$ and $N_{i}$ of the followers of the two ideologies. 
The coefficient characterizing the intensity of the binary conversion  is  $b_{ij}$.
The larger is $b_{ij}$, the more people are converted to the $i$-th ideology. 
In   general  the binary conversion coefficients can be
$b_{ij}=b_{ij}(t,N,N_{\nu},  p_{\mu},C)$. Notice that if $b_{ij}$ is finite, then $b_{ji}=0$, since one converts from $j$ to $i$.
Of course $b_{ii}=b_{jj}=0$: there is no self-conversion.

\item
Another conversion has also to be considered. It is possible that someone can convert to the ideology $i$ simply by interacting with two members of ideology $j$ and $k$. As above one can assume that the relevant term is proportional to the numbers $N_{j}$ and $N_{k}$ of the followers of the two interacting ideologies. 
The coefficient characterizing the intensity of this (still called) binary conversion  is  $t_{ijk}$.
The larger is $t_{ijk}$, the more people are converted to the $i$-th ideology. 
In   general  the relevant binary conversion coefficients can be
$t_{ijk}=t_{ijk}(t,N,N_{\nu}, p_{\mu},C)$. 
Of course $t_{iii}=0$: there is no self-conversion. Also  $t_{iik}$ does not exist, but  $t_{ikk} $ 
 and $t_{ijj} $ can be finite and are factors of $N_{k}^2$ and $N_{j}^2$ respectively. Other coefficients do not exist since one converts  toward $i$.
\end{enumerate}

In general one can have a co-evolution of the environment and the populations with some feedback, i.e. one might have
$p_{\mu} = p_{\mu}(N,N_{\nu},C,t)$, but this  will not be discussed here.

In  the simplest version of the model namely
the case in which all the coefficients 
 are time  and $p_{\mu}$ independent,
 the model system becomes geared by

\begin{equation}\label{mod_syst1}
N=N_{0}+ \sum_{i=1}^{n} N_{i} 
\end{equation}

\begin{equation}\label{mod_syst2}
\frac{dN}{dt} = rN \left(1- \frac{N}{C} \right)
\end{equation}

\begin{equation}\label{mod_syst3}
\frac{dN_{i}}{dt} = r_{i} N_{i} +  \sum\limits_{\begin{array}{*{20}c}
   {j = 0}  \\
   {j \ne i}  \\
\end{array}}^{n} f_{ij} N_{j}+
  \sum\limits_{\begin{array}{*{20}c}
   {j = 0}  \\
   {j \ne i}  \\
\end{array}}^{n}   b_{ij} N_{i} N_{j} 
  +  \sum\limits_{\begin{array}{*{20}c}
   {j = 0}  \\
   {j \ne i}  \\
\end{array}}^{n} \sum\limits_{\begin{array}{*{20}c}
   {k = 0}  \\
   {k \ne i}  \\
\end{array}}^{n} t_{ijk} N_{j} N_{k} 
\end{equation}

\par
A few  remarks are in order.

Notice that arbitrary values are not allowed for the coefficients of the model. They must 
have   values  such that  $N$, $N_{0}$, $N_{1}$, $\dots$, $N_{n}$  
be nonnegative at each $t$.
 
 Terms could apparently be observed to be similar when they refer to the same coupling.  Indeed a $b_{ii} N_{i}^2$ term seem to be of the same nature as $t_{ijj} N_{j} ^2 $. One should not be misled, since the former does not exist, and the latter can never read $t_{jii} N_{i} ^2 $, since it does not pertain to the evolution equation of $N_i$. Nevertheless it is emphasized that the $N_j^2$  term, usually said of higher order, $must$ be conserved in order to maintain correct symmetry conditions.
 
Moreover let us consider the $i$-th population and the binary
conversion characterized by the coefficients $t_{ijk}$ and $t_{ikj}$ where $j$ and $k$
are different from $i$. One could at first think that $t_{ijk} N_{j}N_{k}$
and $t_{ikj} N_{k} N_{j}$ describe one and the same 
process in which the interaction between followers of the $j$-th and $k$-th ideology
leads to a conversion of these to the $i$-th ideology. In general  however one should not 
identify the two terms. In so doing in the general model 
we retain one additional degree of freedom,
i.e., that which allows to distinguish between the ideology that is of the initiator of the
interaction and the ideology of someone who is apparently simply a participant  
in the interaction.

Concerning the content of the $f_{ij}$, $b_{ij}$, and $t_{ijk}$ parameters, one should be aware that they can arise from two ingredients. On one hand the intended final state $i$ maybe attained through imitation and the (positive or negative) influence of neighbours, - along usual majority/minority games, but on the other hand a {\it social value} can also be the cause of a change of ideology, i.e. a shift  toward a {\it more rewarding socially} ideology. The exact form of these coefficients should take such considerations into account in further studies.

Finally, these equations should be completed by two terms if one wants to describe real situations. The first one should be a constant  ($c$) term, as introduced in Eq.(\ref{antoineint}) and a space dependent term of a diffusive type,  $ D_{i} \; \frac{\partial^2 N_{i}}{\partial x^2} $ as  in Eq.(\ref{diffusionD}). The former allows a decay due to say socio-historical circumstances, the latter allows to describe motion of populations in and out the area.

\par
Below we shall consider the dynamics of populations of followers of the ideologies for the
cases of presence of   2  ideologies in  a given area, without diffusion. It will be shown that even without a constant forcing term some intricate situation nevertheless exists.

\subsection{Competition between two ideologies}

In this section  the competition for adepts that the presence of a second 
ideology introduces illustrates the above model. It is shown that an additional ideology leads to  a measurable conflictual tension.
 
Let us consider   the case of two ideologies with populations of
followers $N_{0}$ and $N_{1}$. Let for simplicity all parameters be kept constant. One has

\begin{equation}\label{modN}
\frac{dN}{dt} = r N \left( 1- \frac{N}{C} \right)
\end{equation}

\begin{eqnarray}\label{mod0}
\frac{dN_{0}}{dt}=r_{0} N_{0}  + f_{01} N_{1} + b_{01} N_{0} N_{1} + 
\nonumber\\
 (t_{010}+t_{001}) N_{0} N_{1} +t_{011}  N_{1}^{2} 
\end{eqnarray}
\begin{eqnarray}\label{mod2}
\frac{dN_{1}}{dt}= r_{1} N_{1} + f_{10} N_{0}   + b_{10}N_{0} N_{1} +
 \nonumber\\
(t_{110}+t_{101}) N_{1} N_{0} + t_{100} N_{0}^{2}
\end{eqnarray}

\begin{equation}\label{mod4}
N= N_{0}+ N_{1}  
\end{equation}

The system has a fixed point when only an external field is applied for conversion

\begin{equation} \label{FP}
 \breve{N}_{1} = \frac{C\; f_{10}}{ f_{10}-r_{1}}
\end{equation}

When agent-agent interaction exists, one finds two fixed points, but only one satisfies  $ \breve{N}_{1}\ge 0$. Its value depends on the intial condition, i.e. whether  $ N_1(0)/\breve{N}_{1} \ge$  or $ \le 1$. 
The number of converted agents is  always $N_0=N- \breve{N}_{1} $. 
As a numerical example let $C=1$, $r_{1}=r_{0}=-0.01$ and $f_{10} =0.02$. Then $\breve{N}_{1} /
 \hat{N_{1}} =0.6$, i.e. the number of followers of the ideology 0 decreases
 by 40\%.
 
 Fig. \ref{fig_leeds1GRfits}  shows  a typical result obtained from the numerical investigation of Eqs.(\ref{modN})-(\ref{mod4}).  A  purely inertial growth and its decline are observed for the population of  followers of  
ideology 1, simultaneously with  the  evolution of the population of followers of
ideology 0.  After a maximum of the latter, thus its  decline,   an  inertial growth of 
the number of followers of    ideology 1 takes place again.  This can be usefully compared with the behavior  discussed   by \cite{Abr+03} and  and that of Eq.({\ref{antoineint}), where in the latter case, no competition is assumed.

Even though the time is measured in Monte-Carlo time units, the coefficient is found to be of the reasonable order of magnitude. Indeed anthropologists \cite{steele} reports  a measure of the birth rates of  nomadic hunters and of  agricultural  settlers to be in the range 0.003-0.03/year: ethnographic evidence  suggests than human population can expand at rates of up to 4\% per year when  colonizing new habitat (quotation from \cite{steele}). Thus in the present case, one has a reproduction rate of   0.005, - which  seems to be a good order of magnitude.  An equivalent order of magnitude, but with a negative sign can be taken for $r_1=-0.005$; the value  $f_{10}= 0.001$, chosen for the display,   is somewhat more arbitrary at this time. 

These results can be also compared to a model of telecommunication competition on a network made of nodes (customers)
\cite{legara}, some of them having no mobile phone at first and others belonging to one of several systems.  The conclusion in \cite{legara} allows some interesting reflection by analogy with the cases of languages and religions.  I quote from \cite{legara}: "schemes targeting local cliques within 
the network are more successful at gaining a larger share of the population than those that 
target users randomly at a global scale (e.g., television commercials, print ads, etc.). 
This suggests that success in the competition is dependent not only on the number of 
individuals in the population but also on how they are connected in the network.".  It should be added that  
the network in our above investigation is such that all agents are fully connected with each 
other, i.e. one considers a fully connected graph. This is equivalent to a mean field approximation 
study. Notice that the links are weighted through the $f_{ij}$ and $b_{ijk}$ coefficients.

\begin{figure}
\centering
\includegraphics[height=5cm,width=8cm]{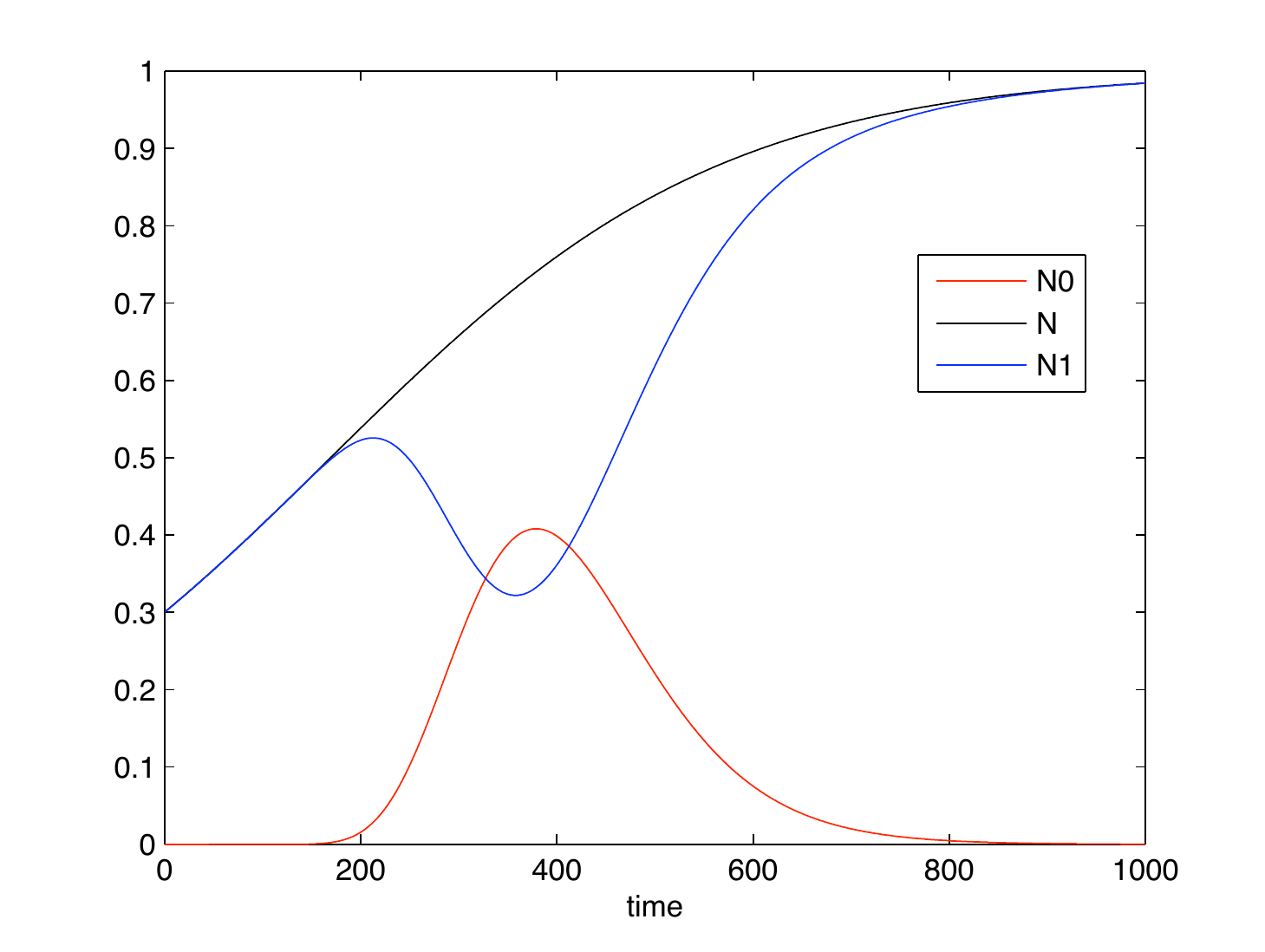}
\caption{\label{fig_leeds1GRfits}  Illustration of the case of two competing  ideologies in populations  $N_0$ and $N_1$,  in an overall  growing ($N$) population: with $r=0.005$, $r_1=-0.005$, and $f_{10}= 0.001$, - thus in the case of no $b$ \& no $t$ influenced  conversion}
\end{figure}

 \subsection{Tension index}
 
 Of course this   causes some tension between the ideologies.  
 A measure of this tension can be found through the index
 
 \begin{equation}\label{strain_meas1}
T_{i;k} = 1 - \frac{N_{i}^{(k)}}{\hat{N}_{i}},
\end{equation}

where $N_{i}^{(k)}$ is the population of the followers of the $i$-th ideology when
the $k$-th ideology is present in the area.
If the ideology is alone then $N_{1}^{(1)} = \hat{N}_{1}$ and the tension index is $T_{1;1}=0$.
If  $N_{1}$ decreases because of the competition with the second ideology,  then
the tension between the ideologies  characterised by  the tension index $T_{i;k}$ increases. 
The above definition for the tension holds even if $N_{1}$ follows some time dependent
trajectory. 

\par
Notice that the tension index can be generalized for the case of an  arbitrary number of ideologies
in the area \cite{vitanov1,vitanov2}. Let $m$ ideologies be present.. 
The tension on the $i$-th ideology in presence of two other 
ideologies, $k$ and $l$ can be defined as

\begin{equation}\label{strain_meas2}
T_{i;k,l} (t) = 1 - \frac{N_{i}^{(k,l)}(t)}{\hat{N}_{i}}
\end{equation}

where $N_{i}^{(k,l)}$ is the population of followers of the $i$-th ideology when the
ideologies $k$ and $l$ operate in the area.
Similarly, the tension on the  $i$-th ideology in presence of three other ideologies, $j$, $k$, $l$  can be measured through

\begin{equation}\label{strain_meas3}
T_{i;j,k,l} (t) = 1 - \frac{N_{i}^{(j,k,l)}(t)}{\hat{N}_{i}}
\end{equation}

where $N_{i}^{(j,k,l)}$ is the number of   followers of the $i$-th ideology
in presence of ideologies $j$, $k$, $l$. In such a way one can define a series of
indices for  the quantification of the tensions among   ideologies competing for
followers.
\par

\section{Discussion and Conclusions}\label{disc_n_concl}

In considering the present studies of physicists on language and/or religion dynamics, one is not only
  interested in their historical course and anthropological features, but also in attempting to unify the description of natural and social phenomena, for the good of humanity progress.

  Here above one has touched upon several aspects, alas neglecting important ones. Mentioned aspects are the size and fraction distribution of denominations in the world. The fluxes are not much studied though. One has insisted on the number of adherents/practitioners, and the role of reproduction rates in the populations, defined though ideologies. The external field influences have been stressed. The conversion through media influence, direct proselytism and indirect, by disgust, conversion have been underlined.  Similarities and differences between religion and languages are numerous. For a physicist, the various time scales are of interest. Same for the basic nucleation-growth problem.  General  differential equations can be written in the spirit of prey-predator systems for describing competitions. The difficulty to introduce quality aspects of language competence and religious practice in simulation through agent based models is an actual challenge. It implies a rethinking of the structure of complex networks.

Many neglected aspects seems to be due to shyness. As the direct comparison of selected word sets of different languages can be used to estimate  their historical distance,  i.e. the measurement of Levenshtein distances between two languages gives an idea of the time of their first common ancestor , one could define and evaluate religion distances, through various historical indicators. Notice that an interesting phase transition-like problem is the evolution from polytheism to monotheism. Finally, the economics of religions has already much attracted researchers \cite{ianna98,PicoliMendes08,antoinism}, but  it could receive more attention from econophysicists. 

 In conclusion, mathematical studies and agent based models on religion and languages are fascinating challenges for  physicists in the future. Combining research in different fields, like anthropology, economy, sociology and so called hard science will remove barriers between fields and thus will allow not only scientific progress but also provide bases for human better understanding of its life in communities.  
 
                \acknowledgments
                
 Thanks to P. Clippe, F. Petroni, S. Pirotte, G. Rotundo,   A. Scharnhorst, D. Stauffer, J.Steele, N. Vitanov  for various critical comments and sometimes joint work on this subject.  Moreover this paper would not have its form nor content without the persuading influence of E. Haven. Thanks also to the   CHESS organizers for some financial support having allowed some presentation of these ideas in Saskatoon in Aug. 2009.

\vskip 1cm


\begin{thebibliography}{99}
\bibitem{footnotetable} The order of items in the Table is somewhat arbitrary but is intended to go from a general, anthropological-like point of view, to a more  statistical physics one. Moreover the list of references is very biased, and does not recognize the enormous amount of works on the subject; surely many of which the author is unaware of, - what he deeply regrets.
\bibitem{footnote2} It is sometimes hard to know or to be sure whether an adherent, a disciple, ... is truly a member of a religious denomination or church. However this caveat pertains to usual problems encountered in sociological investigations.
\bibitem{footnote3} ... admitting indifference, atheism, agnosticism, ... as a sort of religion, from our point of view.
\end{thebibliography}
\end{document}